\documentstyle[twoside,fleqn,epsf,espcrc2]{article}
\newcommand{\ewxy}[2]{\setlength{\epsfxsize}{#2}\epsfbox[10 60 640 570]{#1}}
\newcommand{\del}{{\bf \Delta}}

\newcommand{\delv}{{\bf \del}}
\newcommand{\delfour}{{\Delta^{(4)}}}
\newcommand{\delsq}{\Delta^{(2)}}

\newcommand{\be}{\begin{equation}}
\newcommand{\ee}{\end{equation}}

\newcommand{\mbare}{\mbox{$M_b^0$}}

\newcommand{\nl}{\nonumber \\}

\newcommand{\Ev}{{\bf E}}
\newcommand{\Bv}{{\bf B}}

\newcommand{\sigmav}{\mbox{\boldmath$\sigma$}}

\newcommand{\AmS}{{\protect\the\textfont2
  A\kern-.1667em\lower.5ex\hbox{M}\kern-.125emS}}

\title{New results on heavy hadron spectroscopy with NRQCD}

\author{NRQCD collaboration presented by 
C. T. H. Davies\address{Department of Physics and Astronomy, 
        University of Glasgow, Glasgow G12 8QQ, UK}%
        \thanks{Work done in collaboration with K. Hornbostel, G. P. Lepage, 
                A. Lidsey, P. McCallum, J. Shigemitsu and J. Sloan. Support
                provided by PPARC, DoE, NSF, the Leverhulme Trust and the Fulbright 
                Commission.}}
       
\begin{document}

\begin{abstract}
We present results for the spectrum of $b\overline{b}$ bound states 
in the quenched approximation for three different values of the 
lattice spacing. Results for spin-independent splittings are shown 
to have good scaling behaviour; spin-dependent splittings are
more sensitive to discretisation effects. We discuss what needs to 
be done to match the experimental spectrum.  
\end{abstract}

\maketitle

\section{INTRODUCTION}

Accurate calculations of the hadron spectrum in lattice QCD 
require control of systematic errors, and a major source of 
these errors arises from the use of a space-time 
lattice with finite lattice spacing. Recent understanding of 
these discretisation errors has meant that the `brute 
force' approach of extrapolating to zero lattice spacing has 
been replaced by the requirement of improving the lattice 
action to the point where results at different values of 
$a$ agree to within specified precision. This also allows us 
to use effective actions for QCD for which a continuum extrapolation 
cannot in any case be done, but which are nevertheless good for
particular physics.   

The approach of Non-relativistic QCD fits into this picture
\cite{gpl1,gpl2,oldups}.
The NRQCD action is well-suited to a description of non-relativistic 
quarks such as $b$ quarks within bottomonium or $b$-light mesons
\cite{arifa}. 
In fact we believe that it is the best approach for these
systems. The operators of the action are classified by the 
powers of $v^2/c^2$ that they contain with coefficients that 
can be matched to full QCD in perturbation theory. The 
renormalisability of QCD is lost, however, so an explicit 
momentum cut-off (non-zero lattice spacing) is necessary for 
NRQCD and the coefficients will contain radiative corrections 
which diverge as powers of this cut-off, $\Lambda/m_Q$, so 
$\Lambda/m_Q \rightarrow \infty$ is not possible. Within a range 
of $\Lambda$ around $m_Q$ physical results independent 
of $\Lambda$ are possible, however.
Here we will discuss to what extent this situation 
is realised for current results.     

\subsection{Simulation Details}

We use an evolution equation for quark propagators 
\begin{eqnarray}
  G_{t+1} = 
  \left(1\!-\!\frac{aH_0}{2n}\right)^{n}
 U^\dagger_4
 \left(1\!-\!\frac{aH_0}{2n}\right)^{n} \times \nl
\left(1\!-\!a\delta H\right) G_t .
\label{tevolve}
\end{eqnarray}
$H_0$ is the lowest order (in $v^2/c^2$) term in the Hamiltonian,
the kinetic energy operator:
 \be
 H_0 = - {\delsq\over2\mbare}.
 \ee
The correction terms that we include in 
$\delta H$ are $\cal{O}$$(v^4/c^4)$ \cite{oldups}.
They comprise relativistic corrections to the spin-independent $H_0$ as
well as the first spin-dependent terms that give rise to spin-splittings
in the spectrum. 
 \begin{eqnarray}
\delta H
= &-& c_1 \frac{(\delsq)^2}{8(\mbare)^3} \nl
   &+& c_2 \frac{ig}{8(\mbare)^2}\left(\delv\cdot\Ev -
\Ev\cdot\delv\right) \nl
 &-& c_3 \frac{g}{8(\mbare)^2} \sigmav\cdot(\delv\times\Ev -
\Ev\times\delv) \nl 
 &-& c_4 \frac{g}{2\mbare}\,\sigmav\cdot\Bv  \nl
 &+& c_5 \frac{a^2\delfour}{24\mbare} 
     -  c_6 \frac{a(\delsq)^2}{16n(\mbare)^2} .
\label{deltaH}
\end{eqnarray}
The last two terms in $\delta H$ come from finite lattice spacing 
corrections to the lattice laplacian and the lattice time derivative
respectively ~\cite{gpl2}.  
We tadpole-improve our lattice action by dividing all the 
$U$s that appear by $u_0$, which we take from the fourth root of 
the plaquette ($u_{0P}$). We then work with tree-level values for the 
$c_i$, i.e. 1. 

Table 1 shows the parameters used in the calculation at 3 different values 
of $\beta$, for the standard Wilson gauge action \cite{us-prep}. 

\begin{table}[h]
\begin{center}
\begin{tabular}{ccccc}
\hline
$\beta$ & $aM^0_b$ & $n$ & $u_{0P}$ & $V$ \\
\hline
5.7 & 3.15  & 1 & 0.861 & $12^3 \times 24$ \\
6.0 & 1.71 & 2 & 0.878 & $16^3 \times 32$ \\
6.2 & 1.22 & 3 & 0.885 & $24^3 \times 48$ \\
\hline
\end{tabular}
\end{center}
\caption{ The parameters used. We are 
grateful to the UKQCD collaboration and to Kogut {\it et al} for 
the use of their gauge field configurations. }
\label{table_params}
\end{table}

\section{RESULTS}

\subsection{Radial and orbital splittings}

\begin{figure}[t]
\centerline{\ewxy{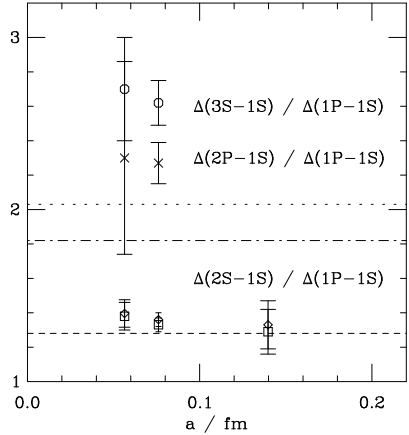}{80mm}}
\caption[hjk]{Various spin-independent ratios of splittings as 
a function of lattice spacing, set by the $\overline{\chi_b} -
\Upsilon$ splitting for NRQCD in the quenched approximation. 
For $\Delta(2S-1S)/\Delta(1P-1S)$ two values are shown. The squares
are the raw results, the diamonds include 
$\cal{O}$$(a^2)$ corections for errors in the Wilson plaquette 
gauge action, estimated from the hyperfine splitting \cite{alpha}. }
\label{siscal}
\end{figure}

Radial and orbital splittings (when spin-averaged to 
remove spin effects) are calculated at next-to-leading 
order both in terms of a non-relativistic expansion and 
in terms of discretisation corrections 
by the action of eq. \ref{deltaH}. We find very little 
remaining $a$ dependence when we take dimensionless 
ratios of spin-independent splittings, as in Figure \ref{siscal}.  
The disagreement with the experimental results, shown as 
lines is presumably an error from the quenched approximation 
since higher order (physical) relativistic corrections 
should have a $\sim$ 1\% effect. This independence from 
the lattice spacing allows us then to extrapolate quenched 
and dynamical ($n_f$ = 2) results to the 
real world \cite{alpha}. 

\subsection{Spin splittings}

The spin splitting that can be calculated most accurately
is the hyperfine splitting between $\Upsilon$ and 
$\eta_b$. No experimental value is known and so the 
lattice results can make a useful prediction if we 
can reduce systematic errors below 10\%. Our aim here 
is to check the scaling of the quenched results before 
attempting an extrapolation in $n_f$
\cite{me-tsukuba}. 

\begin{figure}
\centerline{\ewxy{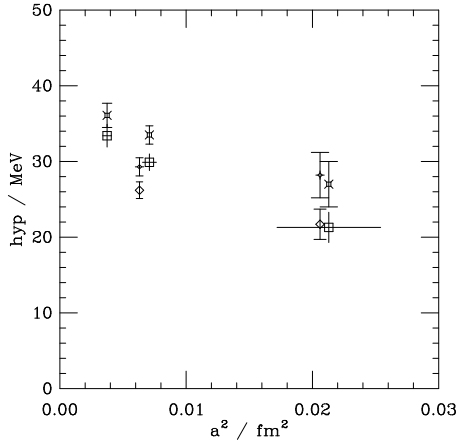}{80mm}}
\caption[hjk]{The hyperfine splitting in physical units vs 
$a^2$ (setting the scale from the 2S-1S splitting) for 
various NRQCD actions in the quenched approximation. 
See text for explanation of the symbols. }
\label{sdscal}
\end{figure}

Because the spin splittings are sensitive to the heavy quark 
mass it is important that this is tuned correctly. 
For our simulation parameters the quark masses we use 
are closest to the correct ones if we fix the scale from 
the 2S-1S splitting. Figure \ref{sdscal} then shows the 
hyperfine splitting in MeV using this scale at the 3 
different lattice spacings. Our results are given by the 
plain squares, and indicate strong scaling violations, as
might be expected of such a short distance quantity with 
$\cal{O}$$(a^2)$ errors. The fancy squares 
are rescaled by $(u_{0P}/u_{0L})^6 $ to give an indication of 
the results that would be obtained by using $u_0$ from 
the Landau gauge link rather than the plaquette. They 
are rather flatter. 

The plain diamonds show results from Manke {\it et al} \cite{manke}
using an action improved to the next order for spin-dependent
terms \cite{gpl2}. This means adding additional relativistic and 
discretisation corrections which however will not change the 
spin-independent spectrum (or lattice spacing). Since 
these calculations use the same parameters as ours, we have 
shown them in Figure \ref{sdscal} using our 2S-1S splitting 
to set the scale. The plain diamonds show much better scaling 
behaviour than the plain squares and illustrate the fact that 
the discretisation and relativistic corrections act in 
different directions \cite{trottier}. The discretisation corrections tend to 
increase the hyperfine splitting (obvious from the 
behaviour of the plain squares) but the relativistic corrections 
tend to reduce the splitting (so that the plain diamonds are 
flatter and below the plain squares).   

The fancy diamonds illustrate approximately what would happen to the plain 
diamonds if $u_{0L}$ had been used instead of $u_{0P}$, using the leading 
order rescaling above. Now very good scaling is observed and this 
is encouraging, but must be checked in a complete calculation 
using $u_{0L}$. Note that the SESAM collaboration have used exactly 
this action with extra spin-dependent corrections and $u_{0L}$ 
in their comparison of results with $n_f$ = 0 and 2 \cite{spitz}. 

Statistical errors in the $p$ fine structure in current calculations 
make it hard to reach clear conclusions about the size of 
scaling violations \cite{us-prep}. 

\section{CONCLUSIONS}

NRQCD actions can give physical results for bottomonium 
splittings which scale accurately for a reasonable range of 
lattice spacings. This is true for spin-independent splittings 
for the action of equation \ref{deltaH}; for spin-dependent splittings 
it is likely that this will happen with the next order of 
relativistic and discretisation corrections \cite{manke,spitz}. 
At this level radiative corrections to the leading order terms 
also have an effect and it will be necessary to decide 
which $u_0$ to use and what the radiative corrections beyond 
tadpole-improvement are
\cite{trott-lat97}.

\end{document}